# On the Future High Energy Colliders

Vladimir SHILTSEV

*Fermilab[1]*

*PO Box 500, Batavia IL, 60510, USA*

High energy particle colliders have been in the forefront of particle physics for more than three decades. At present the near term US, European and international strategies of the particle physics community are centered on full exploitation of the physics potential of the Large Hadron Collider (LHC) through its high-luminosity upgrade (HL-LHC). A number of the next generation collider facilities have been proposed and are currently under consideration for the medium and far-future of accelerator-based high energy physics. In this paper we offer a uniform approach to evaluation of various accelerators based on the feasibility of their energy reach, performance potential and cost range. We briefly review such post-LHC options as linear e+e- colliders in Japan (ILC) or at CERN (CLIC), muon collider, and circular lepton or hadron colliders in China (CepC/SppC) and Europe (FCC-ee and FCC-pp). We conclude by taking a look into ultimate energy reach accelerators based on plasmas and crystals, and discussion on the perspectives for the far future of the accelerator-based particle physics. Extended version of this analysis has recently been presented at the *EPS Conference on High Energy Physics* (22–29 July 2015, Vienna, Austria), see [1].

PRESENTED AT

DPF 2015
The Meeting of the American Physical Society
Division of Particles and Fields
Ann Arbor, Michigan, August 4-8, 2015

# 1 Introduction:

Development of energy frontier colliders over the past five decades initiated a wide range of innovation in accelerator physics and technology which resulted in 100-fold increase in energy (for both hadron and lepton colliding facilities) and $10^4$-$10^6$ fold increase of the luminosity. At the same time, it was obvious that the progress in the maximum c.o.m. energy has drastically slowed down since the early 1990's and the lepton colliders even went backwards in energy to study rare processes – see, e.g., Fig.1 in [2]. Moreover, the number of the colliding beam facilities in operation has dropped from 9 two decades ago to 5 now (2015). In this article we briefly review several future collider options which can be schematically bunched in three groups: *"near future"* facilities with possible construction start within a decade - such as the international *e+e-* linear collider in Japan (ILC) [3] and circular *e+e-* colliders in China (CepC) [4] and Europe (FCC-ee) [5]; *"future"* colliders with construction start envisioned 10-20 years from now – such as linear *e+e-* collider at CERN (CLIC) [6], muon collider [7], and circular hadron colliders in China (SppC) [4] and Europe (FCC-pp) [5]; and an ultimate *"far future"* collider with time horizon beyond the next two decades [2].

Discussion of the options for the future HEP accelerators usually comes to the question of the right balance between the physics reach of the future facilities and their feasibility [2, 8, 9]. Affordable cost of the frontier facility is crucial. As of today, the world's particle physics research budget can be estimated to be roughly 3B$ per year. Under the assumption that such financial situation will not change by much in the future and that not more than 1/3 of the total budget can be dedicated to construction of the next energy frontier collider over approximately a decade, one can estimate the cost of a globally affordable future facility to be about or less than 10B$ (at today's prices).

An analysis of the known costs of large accelerator facilities has been undertaken in [10]. Based on publicly available costs for 17 large accelerators of the past, present and those currently in the planning stage it was shown that the "total project cost (TPC)" (sometimes cited as "the US accounting") of a collider can be broken up into three major parts corresponding to "civil construction", "accelerator components", "site power infrastructure". The three respective cost components can be parameterized by just three parameters – the total length of the facility tunnels $L_f$, the center-of-mass or beam energy $E$, and the total required site power $P$ - and over almost 3 orders of magnitude of $L_f$, 4.5 orders of magnitude of $E$ and more than 2 orders of magnitude of $P$ the so-called "$\alpha\beta\gamma$-cost model" works with ~30% accuracy [10]:

*Total Project Cost* $\approx \alpha \times (Length/10km)^{1/2} + \beta \times (Energy/TeV)^{1/2} + \gamma \times (Power/100MW)^{1/2}$ , (1) where coefficients $\alpha=2B\$/(10\ km)^{1/2}$, $\gamma=2B\$/(100MW)^{1/2}$, and accelerator technology dependent coefficient $\beta$ is equal to *10 B$/TeV$^{1/2}$* for superconducting RF accelerators, *8 B$/TeV$^{1/2}$* for normal-conducting ("warm") RF, *1B$/TeV$^{1/2}$* for normal-conducting magnets and *2B$/TeV$^{1/2}$* for SC magnets (all numbers in 2014 US dollars). Table 1 presents main parameters of the future colliders under discussion and their estimated TPCs. Of note for this discussion is significantly lower cost of doing business in Asia, particularly, in China – for example, comparison of modern synchrotron light sources shows a factor of about 3 lower construction cost for comparable facilities.

Table 1: Main parameters (c.o.m. energy $E_{cm}$, facility size $L_f$, site power $P$) of the collider projects and their estimated total project cost $TPC$ according to the phenomenological $αβγ$–model [10].

|  | $E_{cm}$, TeV | $L_f$, km | $P$, MW | Region | $αβγ$–TPC, $B (est.) |
|---|---|---|---|---|---|
| | | | "Near" Future | | |
| **CepC** | 0.25 | 54 | ~500 | China | 10.2 ±3 |
| **FCC-ee** | 0.25 | 100 | ~300 | CERN | 10.9 ±3 |
| **ILC** | 0.5 | 36 | 163 | Japan | 13.1 ±4 |
| | | | Future | | |
| **CLIC** | 3 | 60 | 589 | CERN | 27.0 ±8 |
| **μμ-Collider** | 6 | ~20 | 230 | US ? | 14.4 ±5 |
| **SppC** | ~50 | 54 | ~300 | China | 25.5 ±8 |
| **FCC-pp** | 100 | 100 | ~400 | CERN | 30.3 ±9 |
| | | | "Far" Future | | |
| **X-Collider** | ≤1000 | ≤10 | ≤100 | ? | ≤ 10 |

## 3  Discussion: Frontier Accelerator HEP Facility Options

All three "near future" colliders are based on well developed technologies of NC magnets and SC RF and from that point of view their abilities to reach the required c.o.m energies ("energy feasibility") have no serious doubts. The feasibility of performance with $L\sim(2\text{-}5)\cdot 10^{34}$ cm$^{-2}$s$^{-1}$ per IP is not fully guaranteed due to a number of challenges, such as extraordinary overall facility power consumption (300-500 MW), heat load due to HOM heating in the cold SC RF cavities and beamstrahlung-limited dynamic aperture for circular $e+e-$ colliders CepC and FCC-ee, and the beam emittance generation and preservation in the main linacs and positron production for the ILC. All three facilities are on the brink of financial feasibility if the latter is defined at the TPC of 10B$ (note, that the publicly announced cost estimate of the ILC in the "European accounting" is 7.8B$ and 13,000 FTE-years of labor [3]).

Among the ("medium") future colliders, only a muon collider is based on the established technology of SC magnets and SR RF and, therefore, can guarantee the energy reach of up to 3-6 TeV c.o.m. It also seems relatively cost-effective and potentially affordable – see Table 1. Unfortunately, at the present, the performance of the muon collider can be assured at the level two to three orders of magnitude below the design luminosity goal of $2\cdot 10^{34}$ cm$^{-2}$s$^{-1}$ and the performance feasibility requires convincing demonstration of the 6-D ionization cooling of muons. The MICE experiment at RAL is expected to provide the first experimental evidence of the muon cooling by 2018. Feasibility of the 3 TeV energy reach of the CLIC collider based on the novel two-beam acceleration in 12 GHz normal conducting RF structures has only recently been demonstrated in a small scale CTF3 test facility where average accelerating gradients of 100 MV/m were achieved with acceptable RF cavity breakdown rates [6]. The luminosity goal of CLIC $L=5\cdot 10^{34}$ cm$^{-2}$s$^{-1}$ is significantly more challenging than that of the ILC, though the design report indicates no principal showstoppers. The biggest issue for CLIC is its enormous site power consuption of about 600 MW and anticipated cost which probably can not be currently considered as affordable – see Table 1. Even a six-times smaller version of a 0.5 TeV c.o.m. $e+e$-collider based on the CLIC technology has been found quite expensive at 7.4-8.3BCHF and 14,100-15,700 FTE-years of labor [6]. Finally, the proton-proton supercolliders such as FCC-hh and SppC can not be claimed as "energy-feasible" as they require development of ~16T SC magnets which are at the edge of the reach of not-yet-fully developed Nb$_3$Sn superconductor technology. Their required luminosity target of above $5\cdot 10^{34}$ cm$^{-2}$s$^{-1}$ is not achievable until critical issues of the synchrotron radiation heat load in the cold magnets, machine protection, ground motion and many others are addressed [11]. The biggest challenge of such huge machines with 60 to 100 km circumferences is their cost. Indeed, according to the $\alpha\beta\gamma$-model Eq.(1), the cost of 100 km long accelerator facility with some 400MW of site power and based on today's SC magnets can be estimated as $TPC=2\times(100/10)^{1/2}+2\times(100\ TeV/1TeV)^{1/2}+2\times(400/100)^{1/2}=30.3B\$\pm 9B\$$. As the biggest share of the TPC is for the magnets, the primary goal of the long-term R&D program should be development of ~16T SC dipole magnets which will be significantly (by a factor 3-5) more cost effective per TeV (or Tesla-meter) then those of, say, LHC.

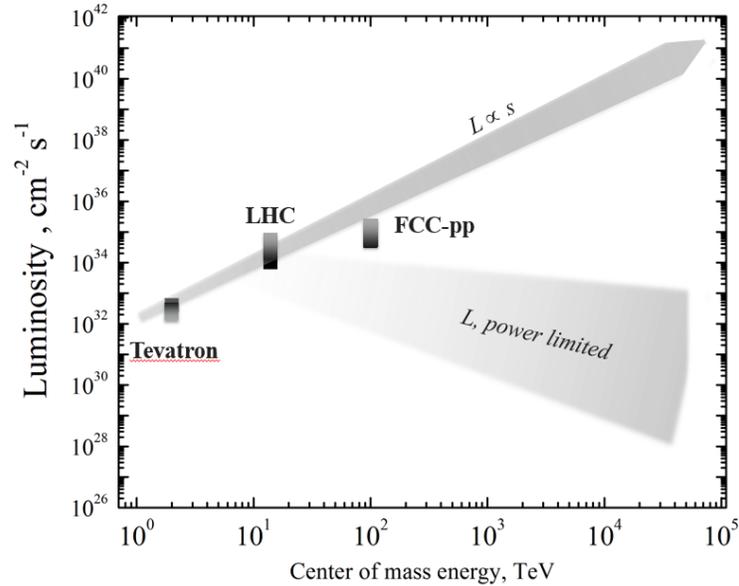

Fig.1: "Luminosity vs Energy" paradigm shift (see text)

While talking about frontier colliders, one should take into account the availability of experts and also the fact that due to extremely complex nature of these accelerators it takes time to get to design luminosity - often as long as 3-7 years [12] – and that should also be taken into account in any realistic plans.

Finally, one can try to assess options for "far future" post-FCC energy frontier collider facility with c.o.m. energies (20-100 times the LHC (300-1000 TeV). We surely know that for the same reason the circular *e+e-* collider energies do not extend beyond the Higgs factory range (~0.25 TeV), there will be no circular proton-proton colliders beyond 100 TeV because of unacceptable synchrotron radiation power – they will have to be linear. It is also appreciated that even in the linear accelerators electrons and positrons become impractical above about 3 TeV due to beam-strahlung (radiation due to interaction at the IPs) and about 10 TeV due to radiation in the focusing channel (<10 TeV). This leaves only *µ+µ-* or *pp* options for the "far future" colliders. If we further limit ourselves to affordable options and request such a flagship machine not to exceed ~10 km in length then we seek a new accelerator technology providing average gradient of >30 GeV/m (compare with $E/L_f$~ 0.5 GeV per meter in the LHC). There is only one such option known now: dense plasma as in, e.g., crystals, that excludes protons because of nuclear interactions and leaves us with muons as the particles of choice [2]. High luminosity can not be expected for such a facility if we limit the beam power and, with necessity, the total facility site power to some affordable level of ~100MW. Indeed, as the energy of the particles *E* grows, the beam current will have to go down at fixed power *I=P/E*, and, consequently, the luminosity will need to go down with energy – see Fig.1. The paradigm shift from the past collider experience when luminosity scaled as $L \sim E^2$ will need to happen in the "far future" of HEP.